\documentclass[preprint,12pt]{elsarticle}

\usepackage{amssymb}
\usepackage{amsmath}
\usepackage{graphicx}

\begin{document}

\begin{frontmatter}



\title{A note on the improved $(G'/G)-$ expansion method}


\author{M. S. Abdel Latif}

\address{Mathematics and Engineering Physics Dept.,
            Faculty of Engineering,\\ Mansoura University, Mansoura,
          Egypt.}

\begin{abstract}
In this paper, we show that the improved $(G'/G)-$ expansion method
is equivalent to the tanh method and gives the same exact solutions
of nonlinear partial differential equations.
\end{abstract}

\begin{keyword}
Exact solution \sep Improved $(G'/G)-$ expansion method  \sep tanh
method

\end{keyword}

\end{frontmatter}


\section{Introduction}
Recently, Many methods have been proposed for obtaining exact
traveling wave solutions of partial differential equations. An
example of these methods is the $(G'/G)-$ expansion
method~\cite{wang} which is used to obtain traveling wave solutions
of many models (see for example \cite{Aslan, Elboree}). Also, many
improved and extended versions of this method have been proposed to
get more exact solutions of partial differential equations (see for
example \cite{Guo1, Guo, Alama}).

Many papers are published to comment on the classical version of the $(G'/G)-$ expansion method. For example, the
equivalence between the $(G'/G)-$ expansion method and the tanh
method is proved in \cite{ping, kudryashov, parkes}. Moreover, in
\cite{aslan}, it is shown that the $(G'/G)-$ expansion method is a
specific form of the simplest equation method \cite{kudryashov1}.

The improved $(G'/G)-$ expansion method \cite{Guo} is used to obtain
new exact solutions of some models~\cite{He, He1}. In this paper, we
show that this improved $(G'/G)$- expansion method is equivalent to
the tanh method and doesn't give any new exact solutions of
nonlinear partial differential equations.

\section{The tanh method \cite{kudryashov}}
In this section, we give the detailed description of the tanh
method. Suppose that a nonlinear evolution equation (NLEE) with
independent variable $u$ and two independent variables $x$ and $t$
is given by
\begin{equation}\label{eq:pde}
   H(u,u_t,u_x,u_{tt},u_{xx},u_{xt},...)=0,
\end{equation}
where, $H$ is a polynomial in $u(x,t)$ and its various partial
derivatives, in which the highest order derivatives and nonlinear
terms are involved. To determine $u$ explicitly, one can follow the
following steps:

Step 1: Use the traveling wave transformation:
\begin{equation}\label{eq:travel}
   u=u(\xi),\quad \xi=x-\nu t,
\end{equation}
where, $\nu$ is a constant to be determined latter. Then, the NLEE
\eqref{eq:pde} is reduced to a nonlinear ordinary differential
equation (NLODE) for $u=u(\xi)$:
\begin{equation}\label{eq:ode}
   H(u,u',u'',u''',...)=0.
\end{equation}

step 2: Suppose that the NLODE \eqref{eq:ode} has the following
solution:
\begin{equation}\label{eq:tanh}
   u=\sum_{i=0}^{n}b_i(\tanh(k(\xi-\xi_0))^i,
\end{equation}
where, $k,b_i(i=0,...,n)$ are constants to be determined later,
$\xi_0$ is an arbitrary constant and $n$ is a positive integer to be
determined in step 3.

Step 3: Determine the positive integer $n$ by balancing the highest
order derivatives and nonlinear terms in Eq. \eqref{eq:ode}.

Step 4: Substituting Eq. \eqref{eq:tanh} into Eq. \eqref{eq:ode} and
equating expressions of different power of $(\tanh(k(\xi-\xi_0))^i$
to zero, we obtain coefficients $b_i$ and the parameter $k$.

Step 5: Substituting $b_i$ and  $k$ into Eq. \eqref{eq:tanh}, we can
obtain the explicit solutions of Eq. \eqref{eq:pde} immediately.

\section{The improved $(G'/G)$- expansion method \cite{Guo}}

In this section, we give the detailed description of the improved
(G'/G) -expansion method. To determine $u$  in Eq. \eqref{eq:pde}
explicitly using the improved $(G'/G)$- expansion method, one can
follow the following five steps:

Step 1: Use the traveling wave transformation \eqref{eq:travel} to
reduce the NLEE \eqref{eq:pde} to the NLODE \eqref{eq:ode}

Step 2: Suppose that the NLODE \eqref{eq:ode} has the following
solution:
\begin{equation}\label{eq:gensol}
   u=\sum_{i=-n}^{n}\frac{a_i\left(G'/G\right)^i}{\left(1+\sigma \left(G'/G\right)\right)^i}=
   \sum_{i=-n}^{n}a_i\left(\frac{\left(G'/G\right)}{1+\sigma
   \left(G'/G\right)}\right)^i,
\end{equation}
where, $\sigma$ and $a_i(i=-n,...,n)$ are constants to be determined
later, $n$ is a positive integer, and $G=G(\xi)$ satisfies the
following second order linear ordinary differential equation(LODE):
\begin{equation}\label{eq:ode1}
   G''+\mu G=0,
\end{equation}
where, $\mu$ is a real constant. The general solutions of Eq.
\eqref{eq:ode1} can be listed as follows. When $\mu<0$, we obtain
the hyperbolic function solution of Eq. \eqref{eq:ode1}
\begin{equation}\label{eq:sol1}
   G=A_1\cosh(\sqrt{-\mu }\xi)+A_2\sinh(\sqrt{-\mu }\xi),
\end{equation}
where, $A_1$ and $A_2$ are arbitrary constants. When $\mu>0$, we
obtain the trigonometric function solution of Eq. \eqref{eq:ode1}
\begin{equation}\label{eq:sol1}
   G=A_1\cos(\sqrt{\mu }\xi)+A_2\sin(\sqrt{\mu }\xi),
\end{equation}
where, $A_1$ and $A_2$ are arbitrary constants. When $\mu=0$, we
obtain the linear solution of Eq. \eqref{eq:ode1}
\begin{equation}\label{eq:sol1}
   G=A_1+A_2 \xi,
\end{equation}
where, $A_1$ and $A_2$ are arbitrary constants

Step 3: Determine the positive integer $n$ by balancing the highest
order derivatives and nonlinear terms in Eq. \eqref{eq:ode}.

Step 4: Substituting \eqref{eq:gensol} along with \eqref{eq:ode1}
into Eq. \eqref{eq:ode1} and then setting all the coefficients of
$\left(G'/G\right)^k, (k=1,2,3,...)$of the resulting systems
numerator to zero, yields a set of over-determined nonlinear
algebraic equations for $\nu, \sigma$ and $a_i(i=-n,...,n)$.

Step 5: Assuming that the constants $\nu, \sigma$ and
$a_i(i=-n,...,n)$ can be obtained by solving the algebraic equations
in Step 4, then substituting these constants and the known general
solutions of Eq. \eqref{eq:ode1} into \eqref{eq:gensol}, we can
obtain the explicit solutions of Eq. \eqref{eq:pde} immediately.
\section{Equivalence of the two methods}
In the second step of the improved $(G'/G)$- expansion method let
$y=\frac{\left(G'/G\right)}{1+\sigma \left(G'/G\right)}$,
Eqs.~\eqref{eq:gensol}, \eqref{eq:ode1} are transformed into
\begin{equation}\label{eq:gensol2}
   u=\sum_{i=-n}^{n} a_i y^i,
\end{equation}
\begin{equation}\label{eq:ode2}
   y'+(1+\mu \sigma^2) y^2-2\sigma \mu y+\mu=0.
\end{equation}
The general solution of  Eq. \eqref{eq:ode2} is given by
\cite{kudryashov}
\begin{equation}\label{eq:ode4sol}
   y=\alpha+\beta \tanh\left(\sqrt{-\mu}
   (\xi-\xi_0)\right),
\end{equation}
where, $\alpha=\frac{\sigma \mu}{1+\mu \sigma^2}$ and
$\beta=\frac{\sqrt{-\mu}}{1+\mu \sigma^2}$.

Substituting solution \eqref{eq:ode4sol} into expansion
\eqref{eq:gensol2} we have
\begin{equation}\label{eq:gensol3}
   u=\sum_{i=-n}^{n} a_i \left(\alpha+\beta \tanh\left(\sqrt{-\mu}
   (\xi-\xi_0)\right)\right)^i=u_1+u_2,
\end{equation}
where,
\begin{equation}\label{eq:gensol4}
   u_1=\sum_{i=0}^{n} a_i \left(\alpha+\beta \tanh\left(\sqrt{-\mu}
   (\xi-\xi_0)\right)\right)^i=\sum_{i=0}^{n} b_i \left(\tanh\left(-\mu
   (\xi-\xi_0)\right)\right)^i,
\end{equation}
\begin{multline}\label{eq:gensol5}
   u_2=\sum_{i=-n}^{-1} a_i \left(\alpha+\beta \tanh\left(\sqrt{-\mu}
   (\xi-\xi_0)\right)\right)^i=\\ \sum_{i=1}^{n} a_{-i} \left(\frac{1}{\alpha+\beta \tanh\left(\sqrt{-\mu}
   (\xi-\xi_0)\right)}\right)^i=\\\sum_{i=1}^{n} a_{-i} \left(\frac{\alpha}{\alpha^2-\beta^2}-\frac{\alpha}{\alpha^2-\beta^2}+\frac{1}{\alpha+\beta \tanh\left(\sqrt{-\mu}
   (\xi-\xi_0)\right)}\right)^i=\\\sum_{i=1}^{n} a_{-i} \left(\frac{\alpha}{\alpha^2-\beta^2}+\frac{\beta}{\beta^2-\alpha^2}\frac{\frac{\beta}{\alpha}+\tanh\left(\sqrt{-\mu}
   (\xi-\xi_0)\right)}{1+\frac{\beta}{\alpha} \tanh\left(\sqrt{-\mu}
   (\xi-\xi_0)\right)}\right)^i=\\\sum_{i=1}^{n} a_{-i} \left(\frac{\alpha}{\alpha^2-\beta^2}+
   \frac{\beta}{\beta^2-\alpha^2}\tanh\left(\sqrt{-\mu}
   (\xi-\xi_0)+z\right)\right)^i,\quad
   z=\tanh^{-1}\frac{\beta}{\alpha},
\end{multline}
therefore, $u_2$ may be rewritten as
\begin{equation}\label{eq:gensol6}
   u_2=\sum_{i=0}^{n} c_{i} \left(
   \tanh\left(\sqrt{-\mu}
   (\xi-\xi_0)+z\right)\right)^i,
\end{equation}
hence,
\begin{equation}\label{eq:gensol7}
  u=u_1+u_2=\sum_{i=0}^{n} b_i \left(\tanh\left(-\mu
   (\xi-\xi_0)\right)\right)^i+\sum_{i=0}^{n} c_{i} \left(
   \tanh\left(\sqrt{-\mu}
   (\xi-\xi_0)+z\right)\right)^i,
\end{equation}
since $\xi_0$ is an arbitrary constant, then we can write
\begin{equation}\label{eq:gensol8}
  u=\sum_{i=0}^{n} d_i \left(\tanh\left(-\mu
   (\xi-\xi_1)\right)\right)^i,
\end{equation}
where, $\xi_1$ is an arbitrary constant. It is now clear that the
two methods will give the same solutions expressed in terms of the
tanh function.

\section{Another proof of equivalence}
In this section we will proof that The solution formula
\eqref{eq:gensol} will give solutions in the form of the tanh
function and the rational function only.

Case 1: When $\mu<0$, we have
\begin{multline}\label{eq:equiv1}
   \frac{G'}{G}=\sqrt{-\mu}\frac{A_2\cosh(\sqrt{-\mu }\xi)+A_1\sinh(\sqrt{-\mu }\xi)}{A_1\cosh(\sqrt{-\mu }\xi)+A_2\sinh(\sqrt{-\mu }\xi)}
=\sqrt{-\mu}\frac{1+\frac{A_1}{A_2}\tanh(\sqrt{-\mu
}\xi)}{\frac{A_1}{A_2}+\tanh(\sqrt{-\mu }\xi)}\\=
\sqrt{-\mu}\coth(\sqrt{-\mu}\xi+d),\quad
d=\tanh^{-1}\frac{A_1}{A_2},
\end{multline}
\begin{multline}\label{eq:equiv2}
\frac{1+\sigma \left(G'/G\right)}{\left(G'/G\right)}=\frac{G+\sigma
G'}{G'}=\frac{G}{G'}+\sigma=\frac{1}{\sqrt{-\mu}}\tanh(\sqrt{-\mu}\xi+d)+\sigma,
\end{multline}

\begin{multline}\label{eq:equiv3}
\sum_{j=-n}^{-1}a_j\left(\frac{\left(G'/G\right)}{1+\sigma
\left(G'/G\right)}\right)^j=\sum_{j=1}^{n}a_{-j}\left(\frac{1+\sigma
\left(G'/G\right)}{\left(G'/G\right)}\right)^j\\
=\sum_{j=1}^{n}a_{-j}\left(\frac{1}{\sqrt{-\mu}}\tanh(\sqrt{-\mu}\xi+d)+\sigma\right)^j=
\sum_{j=0}^{n}b_j(\tanh(\sqrt{-\mu}\xi+d))^j,
\end{multline}

\begin{multline}\label{eq:equiv4}
\frac{\left(G'/G\right)}{1+\sigma \left(G'/G\right)}=\frac{
G'}{G+\sigma
G'}=\frac{\sqrt{-\mu}}{\tanh(\sqrt{-\mu}\xi+d)+\sigma\sqrt{-\mu}}\\
=\left(\frac{\mu \sigma}{1+\mu \sigma^2}-\frac{\mu \sigma}{1+\mu
\sigma^2}+\frac{\sqrt{-\mu}}{\tanh(\sqrt{-\mu}\xi+d)+\sigma\sqrt{-\mu}}\right)\\
=\left(\frac{\mu \sigma}{1+\mu \sigma^2}+\frac{\sqrt{-\mu}}{1+\mu
\sigma^2}\frac{\frac{1}{\sigma
\sqrt{-\mu}}+\tanh(\sqrt{-\mu}\xi+d)}{1+\frac{1}{\sigma
\sqrt{-\mu}}\tanh(\sqrt{-\mu}\xi+d)}\right)\\
=\frac{\mu \sigma}{1+\mu \sigma^2}+\frac{\sqrt{-\mu}}{1+\mu
\sigma^2}\tanh(\sqrt{-\mu}\xi+d+k),\quad k=\tanh^{-1}\frac{1}{\sigma
\sqrt{\mu}},
\end{multline}
\begin{multline}\label{eq:equiv5}
\sum_{j=1}^{n}a_j\left(\frac{\left(G'/G\right)}{1+\sigma
\left(G'/G\right)}\right)^j =\sum_{j=1}^{n}a_{j}\left(\frac{\mu
\sigma}{1+\mu \sigma^2}+\frac{\sqrt{-\mu}}{1+\mu
\sigma^2}\tanh(\sqrt{-\mu}\xi+d+k)\right)^j\\=
\sum_{j=0}^{n}c_j(\tanh(\sqrt{-\mu}\xi+d+k))^j,
\end{multline}
So, in this case (when $\mu<0$) Eq. \eqref{eq:gensol} can be
rewritten as
\begin{multline}\label{eq:equiv6}
u=\sum_{j=-n}^{n}a_j\left(\frac{\left(G'/G\right)}{1+\sigma
\left(G'/G\right)}\right)^j=\sum_{j=0}^{n}b_j(\tanh(\sqrt{-\mu}\xi+d))^j+\\
\sum_{j=0}^{n}c_j(\tanh(\sqrt{-\mu}\xi+d+k))^j=
\sum_{j=0}^{n}e_j(\tanh(\sqrt{-\mu}(\xi-\xi_0))^j,
\end{multline}
where, $\xi_0$ is an arbitrary constant.

Case 2: When $\mu>0$, we have
\begin{multline}\label{eq:equiv7}
   \frac{G'}{G}=\sqrt{\mu}\frac{A_2\cos(\sqrt{\mu }\xi)-A_1\sin(\sqrt{\mu }\xi)}{A_1\cos(\sqrt{\mu }\xi)+A_2\sin(\sqrt{\mu }\xi)}
=\sqrt{\mu}\frac{1-\frac{A_1}{A_2}\tan(\sqrt{\mu
}\xi)}{\frac{A_1}{A_2}+\tan(\sqrt{\mu }\xi)}\\=
\sqrt{\mu}\cot(\sqrt{\mu}\xi+d),\quad d=\tan^{-1}\frac{A_1}{A_2},
\end{multline}
\begin{multline}\label{eq:equiv8}
\frac{1+\sigma \left(G'/G\right)}{\left(G'/G\right)}=\frac{G+\sigma
G'}{G'}=\frac{G}{G'}+\sigma=\frac{1}{\sqrt{\mu}}\tan(\sqrt{\mu}\xi+d)+\sigma,
\end{multline}

\begin{multline}\label{eq:equiv9}
\sum_{j=-n}^{-1}a_j\left(\frac{\left(G'/G\right)}{1+\sigma
\left(G'/G\right)}\right)^j=\sum_{j=1}^{n}a_{-j}\left(\frac{1+\sigma
\left(G'/G\right)}{\left(G'/G\right)}\right)^j\\
=\sum_{j=1}^{n}a_{-j}\left(\frac{1}{\sqrt{\mu}}\tan(\sqrt{\mu}\xi+d)+\sigma\right)^j=
\sum_{j=0}^{n}b_j(\tan(\sqrt{\mu}\xi+d))^j,
\end{multline}

\begin{multline}\label{eq:equiv10}
\frac{\left(G'/G\right)}{1+\sigma \left(G'/G\right)}=\frac{
G'}{G+\sigma
G'}=\frac{\sqrt{\mu}}{\tan(\sqrt{\mu}\xi+d)+\sigma\sqrt{\mu}}\\
=\left(-\frac{\mu \sigma}{1-\mu \sigma^2}+\frac{\mu \sigma}{1-\mu
\sigma^2}+\frac{\sqrt{\mu}}{\tan(\sqrt{\mu}\xi+d)+\sigma\sqrt{\mu}}\right)\\
=\left(-\frac{\mu \sigma}{1-\mu \sigma^2}+\frac{\sqrt{\mu}}{1-\mu
\sigma^2}\frac{\frac{1}{\sigma
\sqrt{\mu}}+\tan(\sqrt{\mu}\xi+d)}{1+\frac{1}{\sigma
\sqrt{\mu}}\tan(\sqrt{\mu}\xi+d)}\right)\\
=-\frac{\mu \sigma}{1+\mu \sigma^2}+\frac{\sqrt{\mu}}{1-\mu
\sigma^2}\tan(\sqrt{\mu}\xi+d-k),\quad k=\tan^{-1}\frac{1}{\sigma
\sqrt{\mu}},
\end{multline}
\begin{multline}\label{eq:equiv11}
\sum_{j=1}^{n}a_j\left(\frac{\left(G'/G\right)}{1+\sigma
\left(G'/G\right)}\right)^j =\sum_{j=1}^{n}a_{j}\left(-\frac{\mu
\sigma}{1-\mu \sigma^2}+\frac{\sqrt{\mu}}{1-\mu
\sigma^2}\tan(\sqrt{\mu}\xi+d-k)\right)^j\\=
\sum_{j=0}^{n}c_j(\tan(\sqrt{\mu}\xi+d-k))^j,
\end{multline}
So, in this case (when $\mu>0$) Eq. \eqref{eq:gensol} can be
rewritten as
\begin{multline}\label{eq:equiv12}
u=\sum_{j=-n}^{n}a_j\left(\frac{\left(G'/G\right)}{1+\sigma
\left(G'/G\right)}\right)^j=\sum_{j=0}^{n}b_j(\tan(\sqrt{\mu}\xi+d))^j
                       \\+ \sum_{j=0}^{n}c_j(\tan(\sqrt{\mu}\xi+d-k))^j=
                      \sum_{j=0}^{n}e_j(\tan(\sqrt{\mu}(\xi-\xi_0)))^j ,
\end{multline}
where, $\xi_0$ is an arbitrary constant. By considering the formula
\cite{kudryashov2}
\begin{equation}\label{eq:formula}
   \tan(i\alpha)=i\tanh(\alpha), \quad i=\sqrt{-1},
\end{equation}
Eq. \eqref{eq:equiv12} can be reformulated as
\begin{multline}\label{eq:equiv5}
u= \sum_{j=0}^{n}e_j(\tan(\sqrt{\mu}(\xi-\xi_0)))^j=
\sum_{j=0}^{n}e_j(-\tan(i\sqrt{-\mu}(\xi-\xi_0)))^j=\\
\sum_{j=0}^{n}e_j(-i\tanh(\sqrt{-\mu}(\xi-\xi_0)))^j=
\sum_{j=0}^{n}f_j(\tanh(\sqrt{-\mu}(\xi-\xi_0)))^j,\quad \mu<0,
\end{multline}
which is equivalent to the solution \eqref{eq:equiv6} in case 1.

Case 3. When $\mu=0$, in this case we will simply obtain the
rational solution.

\section{Conclusion}
It is shown that the improved $(G'/G)-$ expansion method is
equivalent to the tanh method and doesn't give any new exact
solutions of nonlinear  partial differential equations.

\end{document}